\documentstyle[12pt]{article}
\newcommand{\beq}{\begin{equation}}
\newcommand{\eeq}{\end{equation}}
\newcommand{\beqn}{\begin{eqnarray}}
\newcommand{\eeqn}{\end{eqnarray}}
\def\vdir{v\kern-7.8pt\Big{/}}
\def\pdir{p\kern-7.8pt\Big{/}}

\begin{document}

\title{Prediction of beauty particle masses with the heavy quark
effective theory (II)}
\vskip 2.5truecm
\author{
U.~Aglietti\\
SISSA-ISAS, Via Beirut 2, 34014 Trieste, Italy \\
INFN, Sezione di Trieste, Via Valerio 2, 34100 Trieste, Italy\\}
\date{}
\maketitle
\begin{abstract}
\noindent

The effective theory for heavy quarks has additional symmetries with
respect to $QCD$, which relate charm and beauty
hadron masses. Assuming the spectrum of charmed particles,
we predicted in a previous work the masses of some beauty particles.
The predictions of the $\Lambda_b$ mass, $M(\Lambda_b)=5630~MeV$, and of the
$B_s$ mass, $M(B_s)=5379~MeV$, are in agreement with present
experimental data.
We continue this work using recent experimental data on charm
hadron masses. The results are:
$M(\Sigma_b)=5822\pm 6~MeV,~M(\Sigma_b^*)-M(\Sigma_b)=33\pm 3~MeV$,
$M(\Omega_b)=6080\pm 7~MeV,~M(\Omega_b^*)-M(\Omega_b)=32\pm 3~MeV$,
$M(\Lambda_b^*)=5945\pm 3~MeV,~M(\Lambda_b^{**})-M(\Lambda_b^*)=15\pm 1~MeV$.
When experimental data for beauty hadron masses are available,
a comparison with the theoretical values allows a quantitative
estimate of the corrections to the static theory, which contain
informations on hadron dynamics at low energy.

\end{abstract}

\newpage
\section*{~~~~}

The effective theory for heavy quarks \cite{voc5}-\cite{voc21}
has additional symmetries with
respect to $QCD$, called spin-flavor symmetries \cite{voc15}-\cite{voc68}.
These symmetries imply sum rules relating charm and beauty hadron
masses. By means of these sum rules, the masses of some beauty particles
have been predicted in terms of the spectrum of known charmed particles in
ref.\cite{voc1}.
In this note we continue this work using recent experimental data
on charm masses.

The effective lagrangian of a heavy quark $Q$ is given at order $1/M_Q$ by:
\beq
{\cal L} ~=~ Q^{\dagger}(iD_0-M_Q)Q+\frac{1}{2M_Q}Q^{\dagger} \vec{D}^2 Q
+\frac{1}{M_Q}Q^{\dagger} \vec{S}_Q\cdot g\vec{B} Q
\eeq
where $M_Q$ is the heavy quark mass, $\vec{S}_Q$ is the heavy quark spin,
and $\vec{B}$ is the chromomagnetic
field, defined by $B_i=1/2\epsilon_{ijk} F_{jk}$. We consider the operators
of order $1/M_Q$ as
perturbations to the static theory, and we work at first order in perturbation
theory.
The spin $\vec{J}$ of a hadron $H$ is decomposed as:
\beq\label{eq:coupsc}
\vec{J}~=~\vec{S}_Q+\vec{S}_l
\eeq
In the static approximation $\vec{S}_Q$ is conserved; since $\vec{J}$
is conserved, also $\vec{S}_l$ is conserved.
Note that the coupling scheme (\ref{eq:coupsc}) for the angular momenta
is different from that one used in
quark models. In the latter case  one sums the total spin of the quarks
to the total orbital angular momentum.

Then, the mass of a hadron $H$ is given at order $1/M_Q$ by:
\beq
M_H~=~M_Q+\epsilon_l+\frac{1}{2M_Q}\langle\vec{p}^{~2}\rangle_l
    -\frac{1}{M_Q}\langle\vec{S}_Q\cdot\langle g\vec{B}\rangle_l\rangle_J
\eeq
where $\epsilon_l$ is the binding energy of light quarks
in the static approximation
and $\langle ...\rangle_l$ denote averages with light degrees of freedom
in the state $l$.

Since $\vec{S}_l$ is the only vector available,
the matrix element of the chromomagnetic field $\vec{B}$ may be
parameterized as:
\beq
\langle g\vec{B}\rangle_l~=~\mu_l^2~\vec{S}_l
\eeq
where $\mu_l$ is a dimensionful constant of order of the $QCD$ scale
$\Lambda_{QCD}$.

As it stems from eq.(\ref{eq:coupsc}),
the spectrum of hadrons containing a heavy quark and given light flavors
is composed of a series of doublets for $S_l\neq 0$ and of
singlets for $S_l=0$. The spacing between center of gravities of
different doublets is of order $\Lambda_{QCD}$, while the spin-splitting inside
a doublet is of the order of $\Lambda_{QCD}(\Lambda_{QCD}/M_Q)$.

\noindent
The mass of the $\Sigma_c^*$ has been determined recently \cite{simp}:
\beq
M(\Sigma_c^*)~=~2530\pm 5\pm 5~MeV
\eeq
We assume that
$S(\Sigma_c)=1/2$ and $S(\Sigma_c^*)=3/2$, and that these two particles
belong to the same doublet
of the effective theory, with $(S_l)^P=1^+$. We have therefore the sum rule:
\beq\label{eq:cog}
\frac{1}{3}[~M(\Sigma_c)+2M(\Sigma_c^*)~]-M(\Lambda_c)=
\frac{1}{3}[~M(\Sigma_b)+2M(\Sigma_b^*)~]-M(\Lambda_b)
\eeq
Since spin-splitting is produced by $1/M$ terms, we have also the
'scaling' equation:
\beq\label{eq:spsp}
[M(\Sigma _c ^*)-M(\Sigma _c)][M(\Sigma _c)+2M(\Sigma _c ^*)]~=~
[M(\Sigma _b ^*)-M(\Sigma _b)][M(\Sigma _b)+2M(\Sigma _b ^*)]
\eeq
Inserting the experimental masses for the charmed particles and for the
$\Lambda_b$, we have:
\beqn\label{eq:sigmab}
M(\Sigma_b)&=&5822\pm 6~MeV
\\
M(\Sigma_b^*)-M(\Sigma_b)&=&33\pm 3~MeV
\eeqn
where the experimental errors on the masses have been combined in quadrature.

\noindent
Eq.(\ref{eq:sigmab}) has a correction of order $1/M$ of the form:
\beq\label{eq:corr1}
-\frac{1}{2}\left(\frac{1}{M_c}-\frac{1}{M_b}\right)~
(\langle\vec{p}^{~2}\rangle_{\Sigma}-\langle\vec{p}^{~2}\rangle_{\Lambda})
\eeq
It is not easy to estimate the above difference of matrix elements.
This corrections term is produced by spin-spin interactions in
non-relativistic quark models.

The mass of the $\Omega_c$ has been determined recently to be
$M(\Omega_c)\simeq 2715\pm 4~MeV$. We assume that $S^{P}(\Omega_c)=1/2^+$.
The mass of the $\Omega_c^*$ can be predicted from the masses of
$\Sigma_c$ and $\Sigma_c^*$ by means of $SU(3)$ flavor symmetry:
\beq
[M(\Omega_c^*)-M(\Omega_c)][M(\Omega_c)+2M(\Omega_c^*)]=
[M(\Sigma_c^*)-M(\Sigma_c)][M(\Sigma_c)+2M(\Sigma_c^*)]
\eeq
which gives:
\beq
M(\Omega_c^*)-M(\Omega_c)~=~70\pm 6~MeV
\eeq
By means of analogous equations to those in (\ref{eq:cog}) and (\ref{eq:spsp}),
we predict:
\beqn\label{eq:momegab}
M(\Omega_b)&=&6080\pm 7~MeV
\\
M(\Omega_b^*)-M(\Omega_b)&=&32\pm 3~MeV
\eeqn
For the corrections to eq.(\ref{eq:momegab}),
similar considerations hold as those for eq.(\ref{eq:sigmab}).
In the present case, there is an additional correction related to the
breaking of $SU(3)$ flavor symmetry.

Recently, two excited $\Lambda_c$ states have been determined.
The masses of these two resonances, called $\Lambda_c^*$ and $\Lambda_c^{**}$,
have been measured:
\beqn
M(\Lambda_c^*) &=& 2593\pm 2~MeV
\\ \nonumber
M(\Lambda_c^{**}) &=& 2627\pm 1~MeV
\eeqn
Due to the small mass difference, we assume that these states belong to the
same doublet of the effective theory.
The spin and parity have not been determined yet, and can be
predicted with a simple quark model.

Consider a system of three quarks with the same mass, $m_1=m_2=m_3=m$,
interacting by means of elastic forces with constant $k=m\omega^2$
($SU(3)$ flavor symmetry).
This model is solvable and gives a reasonably good description
of the lowest lying excitations \cite{qm}.
The fundamental state $\psi_0(\vec{x}_1,\vec{x}_2,\vec{x}_3)$
has zero orbital angular momentum, $L=0$, positive parity,
$P=+1$, and is symmetric under exchange of quark coordinates $\vec{x}_i$.
The first two excited states $\psi_1$ and $\psi_2$ have $L=1$, negative
parity, $P=-1$, and the same energy $\epsilon=\sqrt{3}\omega$.
$\psi_1$ and $\psi_2$
constitute a two-dimensional representation of the permutation group
acting on $\vec{x}_1$, $\vec{x}_2$ and $\vec{x}_3$.
We send now the mass of one quark to infinity, $m_3\rightarrow\infty$,
without modifying the potential. The infinite mass quark can be placed
in the origin $\vec{x}_3=0$, so that the wavefunctions depend only
on $\vec{x}_1$ and $\vec{x}_2$.
The first two excitations $\psi_1'$ and $\psi_2'$ still have $L=1$ and
$P=-1$. $\psi_1'$ is symmetric under exchange of $\vec{x}_1$ with $\vec{x}_2$
and has energy $\epsilon=\omega$, while
$\psi_2'$ is antisymmetric and has energy $\epsilon=\sqrt{3}\omega$.
Sending the mass of one quark to infinity removes the degeneracy
of the first two excitations.

We identify $\Lambda_c^*$ and $\Lambda_c^{**}$ with the first orbital
excitation $\psi_1'$.
Due to Pauli principle, we have a state of
zero total spin for the light quarks, $S=0$.
Since $\vec{S}_l=\vec{S}+\vec{L}$, we have:
\beq
(S_l)^P~=~1^-
\eeq
Because of eq.(\ref{eq:coupsc}),
$J^{P}(\Lambda_c^*)=1/2^-$ and $J^{P}(\Lambda_c^{**)}=3/2^-$.
Note that we do not use the quark model to predict the values of the masses,
but only to predict the angular momentum of the states when it has not been
measured.

\noindent
With those assumptions, we have the predictions:
\beqn\label{eq:lambdab*}
M(\Lambda_b^*) &=& 5945\pm 3~MeV
\\
M(\Lambda_b^{**})-M(\Lambda_b^{*}) &=& 15\pm 1~MeV
\eeqn
The leading correction to eq.(\ref{eq:lambdab*}) is given by:
\beq\label{eq:gcor}
-\frac{1}{2}\left(\frac{1}{M_c}-\frac{1}{M_b}\right)
(\langle\vec{p}^{~2}\rangle_{\Lambda*}-\langle\vec{p}^{~2}\rangle_{\Lambda})
\eeq
The correction is expected to be large, because the resonances $\Lambda_b^*$
and $\Lambda_b^{**}$ are
orbital excitations of the $\Lambda_b$, and typical momentum
transfers are much greater than in the fundamental state.
The correction (\ref{eq:gcor}) is of the order of $\sim -30~MeV$ in the quark
model considered above.

The dynamics of a heavy quark $Q$ inside a hadron is almost 'frozen',
because typical momentum transfers are of the order of the $QCD$ scale
$\Lambda_{QCD}\ll M_Q$.
Relativistic processes are suppressed. We may take a different point
of view with respect to that one leading to the predictions (\ref{eq:sigmab}),
(\ref{eq:momegab}) and (\ref{eq:lambdab*}).
The heavy quark
acts, in a sense, like a probe  for light quark dynamics.
Comparing the effective theory predictions
with experimental data for beauty hadron masses,
we can evaluate the size of $1/M$
corrections to the static theory. The latter
contain many informations on hadron dynamics at
low energy, like typical momentum transfers, chromomagnetic field strengths,
etc... In general, we expect that $1/M$ corrections increase with the
excitation of the light degrees of freedom
and with increasing light quark masses.

\noindent
The measure of the $\Lambda_b$ mass is:
\beq
M(\Lambda_b)~=~5620\pm 30~MeV,
\eeq
and can be compared with the prediction of the effective theory:
\beq\label{eq:recent}
M(\Lambda_b)~=~5625\pm 2~MeV
\eeq
In ref.\cite{voc1}, the estimate $M(\Lambda_b)=5630~MeV$ has been given
using the first determination of the $B^*$ mass, $M(B^*)=5331.3\pm 4.7~MeV$.
This value is about
seven $MeV$ below the most recent one $M(B^*)=5324.6\pm 2.1~MeV$,
used in eq.(\ref{eq:recent}).
We derive that:
\beq\label{eq:simili}
\mid\langle\vec{p}^{~2}\rangle_{\Lambda}-\langle\vec{p}^{~2}\rangle_M\mid
{}~<~(0.4~GeV)^2
\eeq
where we assumed that $M_c=1.7~GeV$ and $M_b=5.1~GeV$.
The physical meaning of eq.(\ref{eq:simili}) is that
average momentum transfers are very similar in the fundamental
meson and baryon states.

\noindent
The most recent determination of the $B_s$ mass is given by:
\beq
M(B_s)~=~5373.1\pm 4.2~MeV
\eeq
to be compared with the effective theory prediction:
\beq\label{eq:vg}
M(B_s)~=~5374\pm 3~MeV
\eeq
In ref.\cite{voc1}, the value $M(B_s)=5379~MeV$ has been given
using the first measure of the $B^*$ mass.

The correction to the effective theory prediction (\ref{eq:vg})
is controlled in this case by:
\beq
(~\langle\vec{p}^{~2}\rangle_{M(S=1)}-\langle\vec{p}^{~2}\rangle_{M(S=0)}~)
{}~<~(0.15~GeV)^2
\eeq
where $S$ denotes the strangeness of the meson.
\noindent
The difference of the above matrix elements vanishes in the limit of
$SU(3)$ flavor symmetry, and therefore corrections are expected to be very
small. Up to now, there is no disagreement between theory and experiment.
More accurate determinations of $M(B_s)$ may lead to determine a non zero
difference, which gives information about the increase of average
momentum transfer to the heavy quark with increasing light quark mass.

By the same reasoning, we expect the estimate of the
$\Xi_b$ mass given in ref.\cite{voc1} to have a very small correction. The sum
rule
$M(\Xi_b)-M(\Lambda_b)=M(\Xi_c)-M(\Lambda_c)$,
implies $M(\Xi_b)=5809\pm 3~MeV$. The correction is given by:
\beq
-\frac{1}{2}\left(\frac{1}{M_c}-\frac{1}{M_b}\right)~
(\langle\vec{p}^{~2}\rangle_{\Lambda(S=1)}
 -\langle\vec{p}^{~2}\rangle_{\Lambda(S=0)}),
\eeq
and vanishes in the limit of $SU(3)$ flavor symmetry.  We
mean by $\Lambda$ an antisymmetric state in flavor indices.

For excited states, the agreement is not so good, as it is expected from
physical intuition. A preliminary measure of the $B^{**}$ mass gives:
\beq
M(B^{**})~=~5610\pm 40\pm 20~MeV
\eeq
Identifying this state with the resonance $B_1$ ($J^P=1^+$),
we compare with the theoretical value
\beq
M(B_1)~\simeq~5777~MeV,
\eeq
implying a disagreement of at least $100~MeV$.
We derive the following estimate:
\beq
(0.8~GeV)^2~<~
\langle\vec{p}^{~2}\rangle_{M1}-\langle\vec{p}^{~2}\rangle_{M}
{}~<~(1~GeV)^2
\eeq
where $M1$ denotes the first meson excitation, with $(S_l)^P=3/2^+$.

We conclude that the effective theory of E. Eichten and H. Georgi
has a double interest in
the spectroscopy of heavy quark systems. On the one side, it allows
to predict the masses of many beauty particle masses. In some cases,
predictions are expected to be very sound.
On the other hand,
when accurate measurements of beauty hadron masses are available,
the sum rules of the effective theory allow to estimate $1/M$ corrections,
which contain informations on hadron dynamics at low energy.
\vskip .5 truecm
\centerline{\bf Acknowledgment}
\vskip .5 truecm

I wish to thank G. Martinelli for discussions.

\end{document}